# Correlation-induced superconductivity dynamically stabilized and enhanced by laser irradiation

Kota Ido, Takahiro Ohgoe, Masatoshi Imada

Department of Applied Physics, University of Tokyo, 7-3-1 Hongo,
Bunkyo-ku, Tokyo 113-8656, Japan

**Studies on out-of-equilibrium dynamics have paved a way to realize a new state of matter. Especially, superconductor-like properties above room temperatures recently suggested in copper oxides achieved by selectively exciting vibrational phonon modes by laser have inspired studies on an alternative and general strategy to be pursued for high temperature superconductivity. Here, we show that the superconductivity can be enhanced by irradiating laser to correlated electron systems owing to two mechanisms: First, the effective attractive interaction of carriers is enhanced by the dynamical localization mechanism, which drives the system into strong coupling regions. Secondly, the irradiation allows reaching uniform and enhanced superconductivity dynamically stabilized without deteriorating into equilibrium inhomogeneities that suppress superconductivity. The dynamical superconductivity is subject to the Higgs oscillations during and after the irradiation. Our finding shed light on a way to enhance superconductivity that is inaccessible in equilibrium in strongly correlated electron systems.**

**One Sentence Summary: Strong laser irradiation dynamically stabilizes strong-coupling superconductivity without charge inhomogeneities that is inaccessible in equilibrium.**

INTRODUCTION

After the highest superconducting critical temperature $T_c$ has been recorded in cuprate superconductors around 130K decades ago, the highest $T_c$ at ambient pressure has essentially remained unchanged (*1*). One reason would be that the strong effective attraction between electronic carriers required for higher $T_c$ unavoidably drives the tendency for charge inhomogeneities such as the phase separation in a mesoscopic scale and charge ordering (*2–9*). For this reason, materials may not be stably synthesized even if a sufficiently large effective attraction could be reached. In this article, we pursue a way to circumvent this difficulty by utilizing a new axis of control out of equilibrium.

Superconducting-like properties such as gap formations were reported above the room temperature in pump-probe laser measurements with the intension to excite coherent phonons in cuprates (*10, 11*). Such an approach was also applied to alkali-doped fullerides and a similar phenomenon was observed above the critical temperature at equilibrium (*12*).

In the present study, we focus on an alternative possibility ascribed to a mechanism more generic to strongly correlated electron systems. Our strategy is to control electron correlations by using the dynamical localization effect (*13–15*), where the electron hopping is effectively suppressed by the intense alternating-current (AC) electric field at high frequency. The dynamical localization was recently observed experimentally (*15*). We are also inspired by recent finding that the effective attraction of carriers increases from intermediate to strong coupling regions in strongly correlated electron systems (*7*).

By numerically calculating the time evolution of a standard correlated electron model for the cuprates – Hubbard model on square lattice containing the electronic transfer $t_{\text{hop}}$ and the onsite Coulomb repulsion $U$ – we show that the long-range $d$-wave superconducting correlation is substantially enhanced without causing charge inhomogeneity during a strong laser irradiation modeled by a time-dependent vector potential $\boldsymbol{A}(t)$ with the amplitude $A_0$ and frequency $\omega$ (see Materials and Methods).

**RESULTS**

In tight-binding models, the AC laser irradiation causes a dynamical localization. Namely, the transfer integral $t_{\text{hop}}$ is renormalized as

$$t_{\text{hop}} \rightarrow t_{\text{hop}}^{\text{eff}} = t_{\text{hop}} \mathcal{J}_0(A_0), \tag{1}$$

at long time $t \gg 1/\omega$, where $\mathcal{J}_0$ is the zeroth-order Bessel function (*13,14*). If this reduction works in correlated electron systems as well, the effective interaction $U/t_{\text{hop}}^{\text{eff}}$ increases, because $\mathcal{J}_0(A_0 > 0) < \mathcal{J}_0(0) = 1.0$ is satisfied until the first zero, $\mathcal{J}_0(A_0) = 0$, at $A_0 \sim 2.4$ is reached (*14*). The use of reduced hopping amplitude has been recently theoretically proposed for the purpose of enhancing superconductivity in different contexts (16, 17).

To analyze the model, we employ the variational Monte Carlo (VMC) method (*18–21*). In particular, an innovation in the time-dependent VMC method for the fermion systems (*21*) has enabled the long-time nonequilibrium simulation of correlated electron dynamics in the present system (see Materials and Methods).

**Ground state**

Before showing our results for nonequilibrium dynamics, we briefly summarize the results of the ground states for the Hubbard model. Upon doping hole carriers to a half-filled antiferromagnetic Mott insulator, the superconducting order emerges with a dome-like order-parameter amplitude as a function of the hole doping concentration $\delta$ around 0.1 in the intermediate to strong coupling regions (roughly for $U/t_{\text{hop}} > 5$) (*4, 7, 8, 22–24*), in essential agreement with the experiments in the cuprates, which is, however, strongly competing with charge inhomogeneous states such as phase separation and stripe order (*4–9*).

Figure 1 shows physical quantities of uniform and inhomogeneous superconducting states as functions of $U/t_{\text{hop}}$ for the two-dimensional Hubbard model obtained by the VMC method. The detailed definitions of the physical quantities are given in Materials and Methods.

In Fig. 1(A), we see that the long-ranged part of the $d$-wave superconducting correlation function $P_d^\infty$ for the charge uniform states shows a sharp crossover around $U/t_{\text{hop}} \sim 5$ (yellow region). Note that $P_d^\infty$ is basically square of the superconducting order parameter. However, since the energy of the inhomogeneous state is slightly lower than that of the uniform state for

$U/t_{\text{hop}} \gtrsim 5$ (see Fig. 1(B)), the uniform superconducting states are not the true ground states in the strong coupling region (blue arrow in Fig. 1(A)), indicating that the superconductivity is suppressed by charge inhomogeneities in equilibrium ground states. This is naturally understood, because charge inhomogeneous states can be regarded as superconducting regions bridged and separated by non- or suppressed-superconducting areas similarly to the Josephson network. This is consistent with previous numerical results that the phase separation (or charge inhomogeneities) occurs above $U/t_{\text{hop}} \sim 6$ and preempts most of the dome structure of superconductivity for $\delta \lesssim 0.2$ attained under the charge uniform condition (*4,7*).

We add a note of caution on the quantitative aspects. The exact answer is not available even for the simple Hubbard model and the quantitative range of the superconducting dome as well as the charge inhomogeneous region depend on details of the accuracy in numerical methods (*4-9, 21-23*). However, the tight connection between the superconductivity and the charge inhomogeneities is a widely observed tendency (*2-9, 25*) in accordance with the tendency in the experimental observations (*1*). The strong-coupling superconductivity can be achieved only when the effective attractive interaction becomes strong, while too strong static attraction results in the above-mentioned gathering of electrons with charge inhomogeneity, which constitutes a "double-edged sword".

**Dynamically enhanced superconductivity**

Figure 2 shows the dynamics of $P_d^\infty$ induced by strong laser with the scaled frequency $\omega/t_{\text{hop}} = 15$ simulated after the rising time $t_c = 60$. Here, we confirmed that the long-ranged part of the superconducting correlation becomes nearly independent of the distance and reaches an enhanced nonzero constant after the rising time, supporting the emergence of the *d*-wave long-ranged superconducting order (see Materials and Methods). The initial state is a weakly superconducting ground state for $U/t_{\text{hop}} = 5$. This ratio $U/t_{\text{hop}}$ is close to the values for the cuprates estimated by the recent first-principles calculations (*26*). By the laser irradiation, we see that the time-averaged $P_d^\infty$ for $t \geq t_c$ is substantially enhanced from the initial state. The averages are well consistent with those of the charge-uniform excited states for the expected effective interaction, $U/t_{\text{hop}} \sim 5.78$ for $A_0 = 0.75$ and ~ 6.53 for $A_0 = 1.0$ inferred from Eq. (1). See also Figs. S1 and S2 in section A of Supplemental Materials for results of different system sizes.

These results strongly support that this enhancement is caused by an intrinsic effect of the dynamical localization beyond the finite-size artifact. Other physical quantities also consistently indicate that the laser-driven system is well represented by the corresponding uniform excited state at the effective interaction $U/t_{\text{hop}}^{\text{eff}}$ (see section B and Figs. S3, S4 and S5 in Supplementary Materials ).

More importantly, this enhancement is numerically achieved without causing substantial charge inhomogeneities in contrast to the equilibrium ground state. (See insets in Fig. 2.) The enhancement without the inhomogeneity is not very sensitive to the irradiation process and may partly be associated with larger transition elements of the irradiation term between charge homogeneous states than those between homogeneous and inhomogeneous states, in addition to the fast evolution of the pairing. The laser irradiation dynamically stabilizes a charge-uniform state with enhanced superconductivity that is not possible in the equilibrium ground state. We discuss the mechanism of the dynamical stabilization later.

**Higgs Oscillation**

$P_d^\infty$, which is essentially the squared superconducting order parameter, shows long-period oscillations in time in Fig. 2. The oscillations continue after the laser irradiation (see section C and Figs. S7-S9 of Supplementary Materials), signaling the Higgs amplitude oscillation mainly studied in BCS or phonon-mediated superconductors with the *s*-wave symmetry (*27-31*). Depending on the irradiation process, the system may fall into a linear combination of several charge-uniform excited states at the effective $U/t_{\text{hop}}^{\text{eff}}$, which have different superconducting correlations.

Figure 3 shows the optical conductivity of the uniform superconducting state for $U/t_{\text{hop}} = 5.78$ obtained by the VMC method combined with a procedure employed in Ref. (*32*) as detailed in Materials and Methods. A shoulder around $\omega/t_{\text{hop}} \sim 0.09$ coexisting with the delta function at $\omega/t_{\text{hop}} = 0$ smeared by an artificial damping $\eta = 0.04$ is in accord with the inverse oscillation period of the squared superconducting order $P_d^\infty$ for $A_0 = 0.75$ seen in Fig. 2, which is consistent with the interpretation by the gapped Higgs mode.

**DISCUSSION AND SUMMARY**

For intuitive and deep understanding of the dynamical stabilization of the superconductivity, further studies are needed. However, even at this moment, a possible mechanism of the dynamically stabilized superconductivity consistent with our results can be argued in the following way. In the driven electron systems by laser, carriers coupled to electric fields are shaken in the direction of laser irradiation. Since the frequency of the vector potential modulation is larger than the electron hopping $t_{\text{hop}}$, electrons partially follow the modulation and the retardation causes weakening of the stripe charge-density modulation by the motion of electrons from electron-rich to poor regions, in addition to the dynamical localization effect. The shaking also increases the double occupation of electrons especially around the point of peak absolute value of the electric-field oscillation as one can see in Fig. 4. In the equilibrium condition, the ground state has a larger amplitude of stripe modulation, and a smaller double occupation (by an increment of $\Delta D \sim -0.0007$) than the strongly superconducting metastable state at $U/t_{\text{hop}} \sim 6.53$. Since the energy difference between these two states is tiny (at most less than $0.01 t_{\text{hop}}$), the laser irradiation therefore plays the role of a bias field to favor the metastable state over the ground state. In fact, the increment in the double occupation at the peak (bottom or top) of the electric-field oscillation is comparable to the above increments of the superconducting metastable state relative to the stripe ground state in equilibrium. Then the irradiation is expected to dynamically stabilize the uniform and superconducting state, which is only metastable when the laser irradiation is absent.

Next, we discuss realistic possibilities of enhancing superconductivity in the cuprates by utilizing the present mechanism. The most striking enhancement of the superconductivity is expected if $U/t_{\text{hop}}$ of real materials is in the crossover region in equilibrium. However, despite first-principles calculations (*26*), it is not yet clear which region of $U/t_{\text{hop}}$ is adequate for the cuprates when the single-band Hubbard model representation is applied, since the derivations of low-energy effective Hamiltonians call for further improvements such as in the disentanglement procedure of bands (*33, 34*).

In our example of $A_0 = 0.75$ and $\omega/t_{\text{hop}} = 15$, by using realistic values of the lattice constant $a \sim 0.4$nm and $t_{\text{hop}} \sim 0.45$eV, the maximum intensity of the electric field $F_{\text{max}} = A_0\omega$ is required to be $\sim 120$MV/cm which is relatively strong in terms of so far achieved intensity. One way to alleviate the intensity condition would be to use short-width pulses with small frequencies, which contains high-frequency components because of the energy-time uncertainty relation.

We comment on the heating effect possibly induced by the laser irradiation. Contrary to the quench systems where hopping parameter is suddenly changed from the initial value, the laser irradiation has a possibility of continuously heating up systems in general. In the present calculations, however, such heating is negligible within our simulated timescale since the time-averaged energy does not change during the laser irradiation as we see Fig. S3 in section B of Supplementary Materials. This is related with the fact that the single band Hubbard model with $U/t_{\text{hop}} < 10$ does not have electronic density of states at $\omega/t_{\text{hop}} \geq 10$. We should be careful to treat the heating effect if we use laser with small frequencies below $\omega/t_{\text{hop}} \sim 10$. Note that although the time-averaged energy is increased by laser irradiation from around -0.8 to -0.5 in Fig.S3, most of this increase is accounted by the increase in the ground state energy for the system with the effectively reduced hopping, because it causes a substantial increase in the kinetic energy. Therefore, during the laser irradiation, this increased energy has no way to relax.

Aside from technical problems, we need to avoid the heating of the system caused by the dissipation to lattice degrees of freedom. However, this will be delayed and we may be able to see the enhancement before the heating since time scale of phonons in most materials is above picoseconds, which is much slower than that of electrons. In fact, for $t_{\text{hop}} = 0.45$eV, our results show that the enhanced superconductivity is stable during the irradiation from roughly 100fs after the rising time as we see in Fig. 2. In addition, the excitation energy $\Delta E$ from the ground state at the renormalized hopping is only a few K for $t_{\text{hop}} = 0.45$eV as we see in Fig. 1(B) and the averaged energy is conserved during the laser irradiation as we mentioned above (see section B and Fig. S3 in Supplementary Materials). Therefore, the energy will hardly be relaxed to the lattice degrees of freedom during the irradiation. It is an intriguing future issue to examine longer-time sustainability of the superconductivity by tuning repeated switching on and off process of the irradiation to suppress the heating.

Another problem is multi-band structure in real materials missing in the Hubbard model we considered. In our calculations, no optical transition is allowed since the frequency of light is the order of ten as we have mentioned above. However, in the realistic condition, interband transitions may happen due to multi-band structure in materials, which we ignored. It may be important to use the frequency at which the optical transition is not allowed or at least the transition is allowed only between the energies with small density of states. More detailed studies on the heating by laser and the effects of multi-orbital structure are left for future work.

The present approach is also useful to clarify the mechanism of high-$T_c$ superconductivity in equilibrium, which is still an open issue. For instance, we can distinguish whether $U/t_{\text{hop}}$ of real materials such as the cuprates and the iron-based superconductors is in the weak or strong coupling regions from irradiation effects.

Finally, we comment on the possibility of enhancing *s*-wave superconductivity in alkali-doped fullerenes where light-induced superconductivity was proposed similarly to the cuprates

(*12*). In this system, negative Hund's rule coupling arising from Jahn-Teller phonons was proposed to play an important role in the emergence of superconductivity (*35, 36*). It is an intriguing future issue to study whether the reduction of hopping integrals by the dynamical localization and the resultant relative increase of interactions enhances superconductivity.

In summary, in a model of laser irradiation to strongly correlated electron systems, we found dynamically stabilized and enhanced *d*-wave superconducting states that do not suffer from inhomogeneity. It will inspire further studies on high-$T_c$ superconductivity that is not possible in equilibrium.

## MATERIALS AND METHODS
### Model
The Hubbard model is defined by

$$\mathcal{H} = -\sum_{\langle i,j\rangle,\sigma} t_{\text{hop}} c_{i\sigma}^\dagger c_{j\sigma} + U \sum_{i}^{N_s} n_{i\uparrow} n_{i\downarrow}, \tag{2}$$

where $t_{\text{hop}}$ and $U$ represent the hopping amplitude of electrons between nearest-neighbor sites and the on-site interaction, respectively. The creation and annihilation operators of electrons at site *i* with spin $\sigma$ are expressed as $c_{i\sigma}^\dagger$ and $c_{i\sigma}$, respectively. In the presence of a vector potential, the hopping amplitude is modified as $t_{\text{hop}} \to t_{\text{hop}} \exp[i\boldsymbol{A}(t)\cdot(\boldsymbol{r}_i - \boldsymbol{r}_j)]$, where $\boldsymbol{A}(t)$ is a vector potential at time *t* and $\boldsymbol{r}_i$ is the position vector at site *i*. We consider a square lattice and assume that the vector potential is polarized along the direction $\boldsymbol{d} = (1,1)$, *i.e.* $\boldsymbol{A}(t) = A(t)\boldsymbol{d}$. The time dependence of the vector potential $A(t)$ is taken as

$$A(t) = A_0 g(t) \cos[\omega(t - t_c)], \tag{3}$$

$$g(t) = \begin{cases} \exp\left[-\dfrac{(t - t_c)^2}{2 t_d^2}\right] & (0 \le t \le t_c), \\ 1 & (t_c \le t). \end{cases} \tag{4}$$

We set $t_c = 3 t_d$ and use $t_{\text{hop}}$ and $t_{\text{hop}}^{-1}$ as units of energy and time, respectively.

### Definitions of physical quantities
To study physical properties in equilibrium and nonequilibrium, we measure the energy per site $E/N_s$, its average over an AC cycle $\tilde{E}/N_s$, double occupancy $D$, amplitude of charge density $\Delta n$, charge structure factor $N(\boldsymbol{q})$ and the $d_{x^2-y^2}$-wave superconducting correlations $P_d(\boldsymbol{r})$ defined as

$$E = \langle \mathcal{H} \rangle, \quad \tilde{E}(t) = \frac{1}{2\pi/\omega} \int_t^{t+2\pi/\omega} dt' E(t'), \tag{5}$$

$$D = \frac{1}{N_s} \sum_i \langle n_{i\uparrow} n_{i\downarrow} \rangle, \tag{6}$$

$$\Delta n = n_{\max} - n_{\min}, \tag{7}$$

$$N(\boldsymbol{q}) = \frac{1}{N_s} \sum_{i,j} e^{-i\boldsymbol{q}\cdot(\boldsymbol{r}_i - \boldsymbol{r}_j)} \langle (n_i - n)(n_j - n) \rangle, \tag{8}$$

$$P_d(\boldsymbol{r}) = \frac{1}{2N_s} \sum_i [\langle \Delta_d^\dagger(\boldsymbol{r}_i) \Delta_d(\boldsymbol{r}_i + \boldsymbol{r}) \rangle + \langle \Delta_d(\boldsymbol{r}_i) \Delta_d^\dagger(\boldsymbol{r}_i + \boldsymbol{r}) \rangle], \tag{9}$$

respectively. Here, $n_{\max}(n_{\min})$ are those among $\{\langle n_i \rangle, i = 0, \cdots, N_s\}$ and $n = N/N_s$, where $N$ is the number of electrons. $\langle Q \rangle$ represents the expectation value of an operator $Q$, i.e., $\langle Q \rangle = \frac{\langle \psi | Q | \psi \rangle}{\langle \psi | \psi \rangle}$ where $|\psi\rangle$ represents a wave function. $\Delta_d(\boldsymbol{r}_i)$ represents the $d_{x^2-y^2}$-wave superconducting order parameter defined as

$$\Delta_d(\boldsymbol{r}_i) = \frac{1}{\sqrt{2}} \sum_j f_d(\boldsymbol{r}_j - \boldsymbol{r}_i)(c_{i\uparrow} c_{j\downarrow} - c_{i\downarrow} c_{j\uparrow}), \tag{10}$$

where

$$f_d(\boldsymbol{r}) = \delta_{r_y,0}(\delta_{r_x,1} + \delta_{r_x,-1}) - \delta_{r_x,0}(\delta_{r_y,1} + \delta_{r_y,-1}), \tag{11}$$

is the form factor which describes the $d_{x^2-y^2}$-wave symmetry, $\boldsymbol{r} = (r_x, r_y)$, and $\delta_{k,l} = 1$ for $k = l$ and 0 otherwise is the Kronecker's delta. We also define the averaged value of long-range parts of the superconducting correlation as

$$P_d^\infty = \frac{1}{M} \sum_{r=|\boldsymbol{r}|>5} |P_d(\boldsymbol{r})|, \tag{12}$$

where $M$ is the number of vectors satisfying $r = |\boldsymbol{r}| > 5$. In our calculated data, $P_d(\boldsymbol{r})$ saturates to a nonzero constant if the superconducting order exists (see Fig.S6 in Supplementary Materials).

**Variational Monte Carlo method for correlated electron systems**

We use a variational Monte Carlo method to obtain nonequilibrium states as well as ground states in strongly correlated electron systems (*18-21*). As with the VMC method for ground states, the recent developed time-dependent VMC method for fermion systems provides us with an accurate and efficient way to study nonequilibrium dynamics of correlated electrons in finite-dimensional systems without the negative sign problem (*21*). Thanks to this method, we are able to simulate long-time electron dynamics for two-dimensional carrier-doped correlated systems, whose sizes can be taken sufficiently large to estimate the stabilities of various competing long-range orders on equal footing.

As a trial wave function, we adopt the generalized pair product wave function with correlation factors $|\psi\rangle = \mathcal{P}|\phi\rangle$, where $|\phi\rangle = \left( \sum_{i,j}^{N_s} f_{ij} c_{i\uparrow}^\dagger c_{j\downarrow}^\dagger \right)^{N/2} |0\rangle$ and $\mathcal{P}$ is the correlation factor. Here, $f_{ij}$ are site-dependent variational parameters as noted below. The correlation factor consists of the Gutziwiller, Jastrow and doublon-holon binding factors which are defined as $\mathcal{P}_G = \exp(-g \sum_i n_{i\uparrow} n_{i\downarrow})$, $\mathcal{P}_J = \exp\left(-\sum_{i,j} v_{ij} n_i n_j\right)$ and $\mathcal{P}_{d-h}^{ex} = \exp\left(-\sum_{m=0}^{5} \sum_{l=1,2} a_{(m)}^{(l)} \sum_i \xi_{i(m)}^{(l)}\right)$, respectively. $\xi_{i(m)}^{(l)}$ is 1 when a doublon (holon) exists at the $i$-th site and $m$ holons (doublons) surround at the $l$-th nearest neighbor. Otherwise, $\xi_{i(m)}^{(l)}$ is 0. In this study, we treat $g$, $v_{ij}$, $a_{(m)}^{(l)}$ and $f_{ij}$ as variational parameters. Since many variational parameters are introduced not only to correlation factors but also to the one body part, this type

of trial wave functions can describe quantum states flexibly and accurately in correlated electron systems (see Refs. (*7,19,21*) and Fig. S2). We note that our trial wave function well reproduces the exact results for the time evolution of physical quantities such as energy, momentum distribution, spin structure factor and superconducting correlations in the Hubbard model (*21*). All the variational parameters are optimized simultaneously by using the stochastic reconfiguration (SR) method for imaginary-time or real-time evolution based on the time-dependent variational principle (*18-21, 38-41*) as described later in detail.

In actual calculations, we employ a trial wave function that has a $8 \times 2$ sublattice structure by introducing $8 \times 2 \times N_s$ variational parameters for $f_{ij}$, because it enables flexibly describing realistic and possible inhomogeneous states, *i.e.*, the inhomogeneous state is allowed to emerge dynamically. Note that, since uniform-charge states do not exist for the equilibrium ground state above $U/t_{\text{hop}} \sim 6$, we impose $2 \times 2$ sublattice structure on a trial wave function for $U/t_{\text{hop}} \gtrsim 6$ to examine the charge-uniform lowest excited states.

**Real-time and imaginary-time evolution in variational Monte Carlo method**

In our VMC calculations, ground states and nonequilibrium states are obtained by using the SR method (*18-21*). This method can be regarded as the imaginary-time or real-time evolution of a trial wave function in the truncated Hilbert space based on the time-dependent variational principle (*38-41*). In the case of real-time evolution, by minimizing a distance between $i \frac{d}{dt} |\psi\rangle$ and $\mathcal{H}|\psi\rangle$ in the time-dependent Schrödinger equation in accordance with the time-dependent variational principle (*38-41*), the equation of motion for variational parameters $\boldsymbol{\alpha} = \{\alpha_k | k = 1, \cdots, N_p\}$ are obtained as

$$\dot{\alpha}_k = \frac{\partial \alpha_k}{\partial t} = -i \sum_{l=1}^{N_p} (S^{-1})_{kl} g_l, \quad (13)$$

where $S_{kl} = \langle \frac{\partial}{\partial \alpha_k^*} \frac{\partial}{\partial \alpha_l} \rangle$ and $g_k = \langle \frac{\partial}{\partial \alpha_k^*} \mathcal{H} \rangle - \langle \frac{\partial}{\partial \alpha_k^*} \rangle \langle \mathcal{H} \rangle$ (*20, 21, 39, 41*). In this study, we use the fourth-order Runge-Kutta method to solve Eq. (13). The imaginary-time evolution of variational parameters is obtained by replacing the real-time variable *t* in Eq. (13) with the imaginary-time $\tau = it$. If we solve Eq. (13) for the imaginary-time evolution by using the Euler method, it is nothing but the conventional SR scheme for ground states (*18,19*). In the VMC method, the quantities $S_{kl}$ and $g_k$ are estimated by the Markov-chain Monte Carlo method.

**Optical conductivity**

We also calculate the optical conductivity measurable by a pump-probe experiment by using a laser with weak intensity (*32*). The details of the method are as follows. First, we obtain the time-dependent trial wave function in the system under a laser irradiation with weak intensity $A_{\text{probe}}$. Here, $A_{\text{probe}}$ represents the probe laser polarized along the *x*-direction defined as

$$A_{\text{probe}}(t) = A_0 \exp\left[-\frac{(t-t_c)^2}{2t_d^2}\right], \quad (14)$$

Next, we calculate the expectation value of the current operator defined by

$$J_x(t) = -\langle \frac{\partial \mathcal{H}}{\partial A_{\text{probe}}} \rangle = i \sum_{j,\sigma} \sum_{\delta=\pm \hat{x}} t_{j,j+\delta} e^{i A_{\text{probe}} a \delta} a \delta \langle c_{j\sigma}^\dagger c_{j+\delta,\sigma} \rangle, \quad (15)$$

where $a$ is the lattice constant. To obtain the optical conductivity in Fig. 3, we set $t_c = 0.05$, $t_d = 0.01$ and $A_0 = 0.05$. Finally, we calculate the optical conductivity $\sigma_{xx}$ defined as

$$\sigma_{xx}(\omega) = \frac{J_x(\omega)}{N_s F(\omega)} = \frac{-J_x(\omega)}{N_s(i\omega + \eta)A_{\text{probe}}(\omega)}, \tag{16}$$

where the electric field is defined as $F(t) = -\frac{dA_{\text{probe}}(t)}{dt}$. Note that the Fourier transformations in Eq. (16) is defined by

$$F(\omega) = \int_0^T dt F(t) e^{i(\omega + i\eta)t}, \tag{17}$$

where we set sufficiently large $T$ and small $\eta$ as $T = 200$ and $\eta = 0.04$.

42. Our exact calculations were performed by using the code based on the open-source software HΦ. See arXiv:1703.03637 or https://github.com/QLMS/HPhi.

**Acknowledgments:** Our calculations were performed by using the code based on the open-source software mVMC (*35*).

**Funding:** The authors thank the Supercomputer Center, the Institute for Solid State Physics, the University of Tokyo for the facilities. This work was financially supported by Japan Society for the Promotion of Science through Program for Leading Graduate Schools (MERIT), the MEXT HPCI Strategic Programs for Innovative Research (SPIRE), the Computational Materials Science Initiative (CMSI) and Creation of New Functional Devices and High-Performance Materials to Support Next-Generation Industries (CDMSI). We thank the computational resources of the K computer provided by the RIKEN Advanced Institute for Computational Science through the HPCI System Research project (under the project number hp130007, hp140215, hp150211, hp160201, and hp170263). This work was also supported by a Grant-in-Aid for Scientific Research (No. 22104010, No. 22340090 and No. 16H06345) from Ministry of Education, Culture, Sports, Science and Technology, Japan.

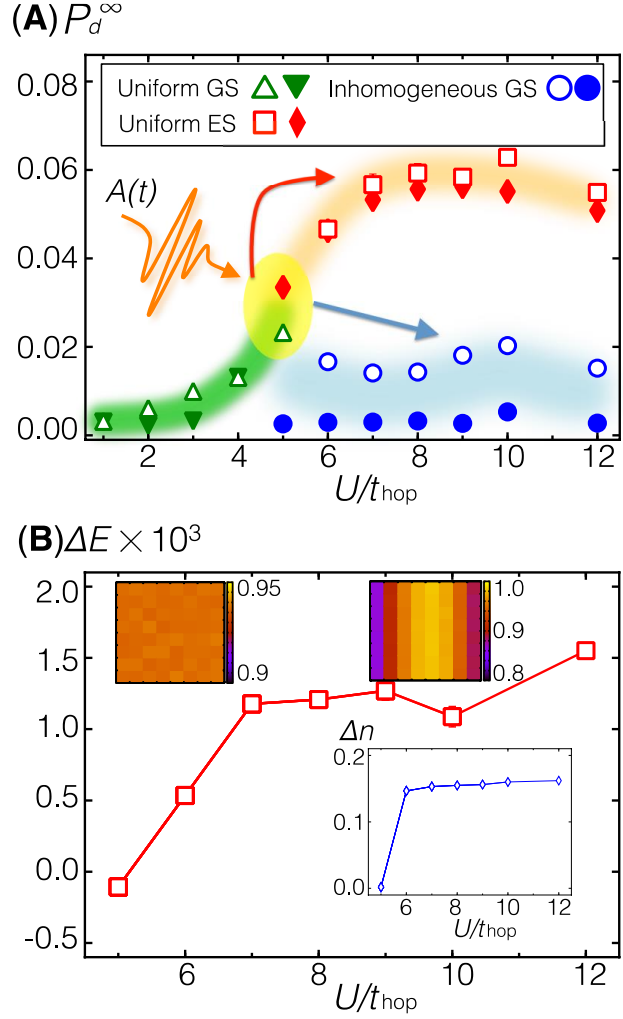

**FIG. 1. Interaction dependence of physical quantities for superconducting states at zero-temperature equilibrium.** (A) Long-range average of the superconducting correlation $P_d^\infty$ in two-dimensional Hubbard model. Open and solid symbols represent the results for size $L \times L = 8 \times 8, \delta \approx 0.06$ and $16 \times 16, \delta \approx 0.1$, respectively. Green triangles (upside-down triangles) and blue circles represent the results of charge-uniform and inhomogeneous ground states (GS), respectively. Red square (diamond) symbols represent the results for charge-uniform metastable excited states (ES). Shaded regions are guides for the eyes. Blue and red arrows represent equilibrium path and hypothetical nonequilibrium process toward uniform excited states with enhanced superconductivity by the laser irradiation $A(t)$, respectively. Error bars indicate the statistical errors arising from the Monte Carlo sampling, but most of them are smaller than the symbol sizes here and in the following figures. (B) Energy difference per site between the uniform state with strong superconductivity and the inhomogeneous state with the weak superconductivity for $L = 8$ and $\delta \approx 0.06$. Insets represent the charge distribution $\langle n_i \rangle$ of the uniform state for $U/t_{\text{hop}} = 5$ (left top), that of the inhomogeneous state for $U/t_{\text{hop}} = 10$ (right top) and the amplitude of the density inhomogeneity (right bottom), respectively.

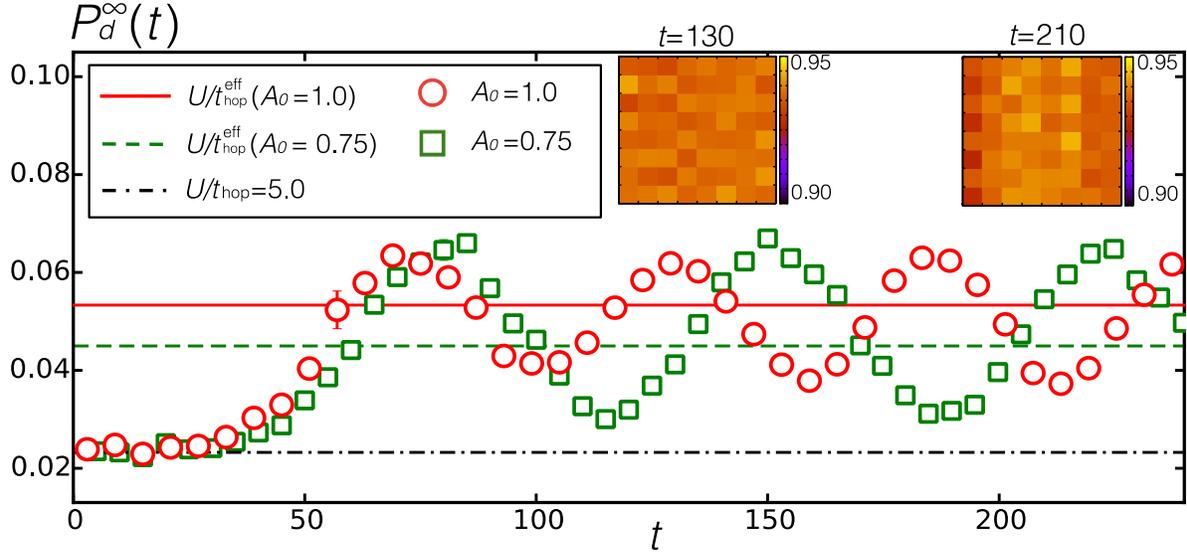

**FIG. 2**. **Time-dependence of $P_d^\infty$ induced by strong laser in two-dimensional Hubbard model.** Parameter set is $(\delta, L, U/t_{\text{hop}}, \omega/t_{\text{hop}}, t_c) = (0.0625, 8, 5.0, 15, 60)$ on an $8 \times 8$ lattice. Red circles and green squares represent the results for $A_0 = 0.75$ and $A_0 = 1.0$, respectively. Red solid and broken green lines indicate corresponding values of $P_d^\infty$ obtained from VMC for the charge-uniform superconducting states by assuming the effective enhancement of $U/t_{\text{hop}}$. Black dash dotted line is the equilibrium value in the absence of the laser irradiation. Insets represent the charge distributions $\langle n(\boldsymbol{r}) \rangle$ for $A_0 = 1.0$ at $t = 130$ and $t = 210$, corresponding to a peak and a bottom of the oscillation, respectively.

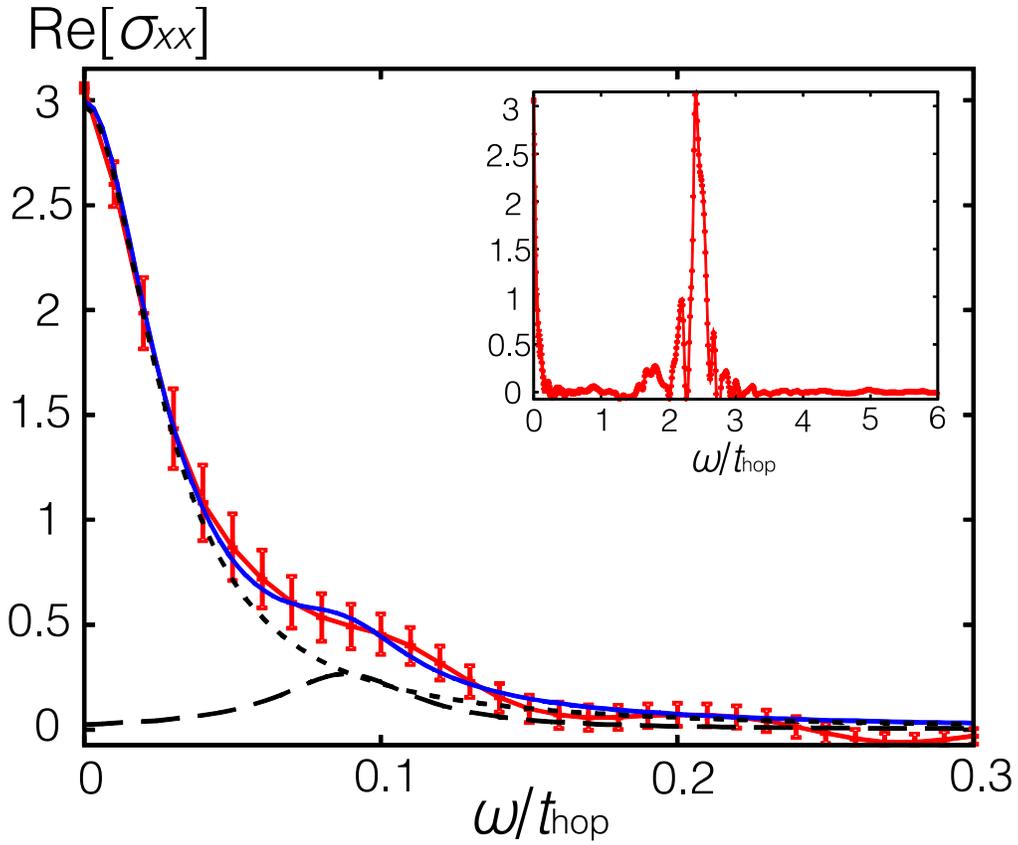

**FIG. 3. Real part of optical conductivity of superconducting state.** Parameters are $L = 8$, $\delta = 0.06$ and $U/t_{\mathrm{hop}} = 5.78$. The results (red curves with symbols) are obtained by using the VMC method combined with a numerical procedure employed in Ref. (*32*). Error bars indicate the statistical errors arising from the Monte Carlo sampling. The inset shows a wider range of $\omega$ dependence. The blue solid line is a fitting obtained by a sum of two Lorentzian curves (dashed and dotted lines).

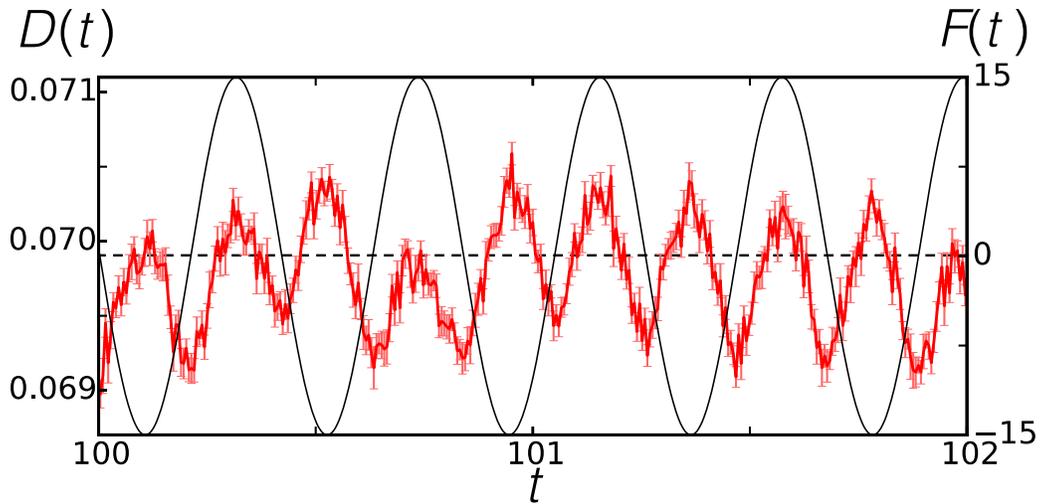

**FIG. 4. Time-dependence of double occupation in short time scale around *t*=100 for the same case as Fig. 2 for $A_0 = 1.0$.** The black curve shows time evolution of electric field $F(t) = -\frac{dA(t)}{dt}$. Dotted line represents the zero line of the electric field.

# Supplementary Materials for "Correlation-induced superconductivity dynamically stabilized and enhanced by laser irradiation"

Kota Ido, Takahiro Ohgoe, Masatoshi Imada

## A. Time-evolved superconducting correlations by laser irradiation in exact calculations

Based on the exact numerical results for small systems, we discuss the possibility of enhancement of superconductivity by the strong laser irradiation (*42*). To obtain the exact results, we directly calculate the formal solution of the time-dependent Schrödinger equation in the full Hilbert space as

$$|\psi(t+\Delta t)\rangle = e^{-i\mathcal{H}\Delta t}|\psi(t)\rangle = \sum_{n=0}^{M-1}\left(-i\mathcal{H}\frac{\Delta t}{n!}\right)^n |\psi(t)\rangle + \mathcal{O}(\Delta t^M). \tag{S1}$$

In Fig. S1 (A), we show the $U/t_{\text{hop}}$-dependence of $P_d(|\mathbf{r}|=r_{\max})$ for ground states. As the parameter set for the system, we choose $(\delta, L, r_{\max}) = (0.25, 4, 2\sqrt{2})$. Here, $r_{\max}$ represents the farthest site from the origin. As seen in this figure, if we consider the case of $U/t_{\text{hop}}=1$ as an initial state and then $t_{\text{hop}}$ is reduced. $P_d(|\mathbf{r}|=r_{\max})$ increases due to the increases in $U/t_{\text{hop}}$. Then if the reduction in the effective hopping is effective by the dynamical localization mechanism through Eq. (1) even in the correlated electron systems, the superconducting correlation may be enhanced.

Figure S1 (B) shows the dynamics of $P_d(|\mathbf{r}|=r_{\max})$ for the Hubbard model in the presence of the laser irradiation. For the laser, we consider the two cases of $(A_0, t_c, \omega) = (1.5, 15, 15)$ and $(A_0, t_c, \omega) = (1.95, 15, 15)$. The initial state is the ground state for $U/t_{\text{hop}}=1$. Within the single-particle picture, the effective on-site interaction is estimated from Eq. (1) as $U/t_{\text{hop}}^{\text{eff}} \approx 1.95$ for $A_0 = 1.5$ and $U/t_{\text{hop}}^{\text{eff}} \approx 3.96$ for $A_0 = 1.95$ for $t \geq t_c$, respectively as shown by arrows in Fig. S1 (A). As seen in Fig. S1 (B), $P_d(|\mathbf{r}|=r_{\max})$ grows for $t < t_c$ and oscillates for $t \geq t_c$ in both cases. The averaged value during $t \geq t_c$ well reproduces the values of the ground state for $U/t_{\text{hop}}^{\text{eff}}$ as expected. These results are consistent with those of VMC calculations in the main text.

## B. Stationarity and spatial saturation of physical quantities

Figure S3 shows the time dependence of the energy $E/N_s$ and its average over an AC cycle $\tilde{E}/N_s$ in the Hubbard model induced by strong laser for $A_0 = 1.0, t_c = 60$ and $\omega/t_{\text{hop}} = 15$. The definitions of these physical quantities are given in Eq. (S4). In Fig. S3, the averaged energy $\tilde{E}/N_s$ is constant during $t > t_c$. This implies that the nonequilibrium state reaches the steady state in $t > t_c$. We also show the dynamics of the double occupancy $D$ and the peak value of the charge structure factor $N(\mathbf{q}_{\text{peak}} = (\pm 2\pi/L, 0))$ in Figs. S4 and S5, respectively. As with the dynamics of the long-range part of the superconducting correlation in Fig. 2, these physical quantities also indicate that the laser-driven system is well represented by the corresponding charge-uniform excited state at the effective interaction $U/t_{\text{hop}}^{\text{eff}}$. In addition, we can clearly see that charge inhomogeneity does not grow up to that for the corresponding inhomogeneous ground state.

In order to discuss the long-range order of superconductivity, we need to see whether the superconducting correlation is saturated at long distances. We show examples of the superconducting correlation in Figs. S2 and S6. For the small system with $L = 4$, we do not observe the saturation of the superconducting correlation (see Fig. S2). Figure S6 shows the dynamics of the superconducting correlations $P_d(r)$ for the $8 \times 8$ lattice. Other parameters are the same as those in Fig. S3. We can see that the long-range part of $P_d(r)$ is saturated to a non-zero value at any time. Thus, the nonequilibrium states we obtained in Fig. 2 can be regarded practically as the long-range ordered superconducting states within our simulation sizes.

## C. Nonequilibrium dynamics of physical quantities after laser irradiation

Figures S7 and S8 show the dynamics of physical quantities during and after strong laser irradiation for $A_0 = 1.0$, $\omega/t_{\text{hop}} = 15$ and $t_c = 60$. The intensity of the laser start to fall at $t = 200$ and we set the falling time $t_f = 60$. The envelop function of the laser $g(t)$ in Eq. (S3) is shown in Fig. S9. In Fig. S7, we see that $P_d^\infty$ decreases as the intensity of the electric field becomes weak because $U/t_{\text{hop}}^{\text{eff}}$ decreases through Eq. (1). We also see that the averaged value of $P_d^\infty$ after the irradiation decreases. This is because the system acquires the additional energy from the laser irradiation during the rising and falling time regions (see Fig. S8). However, it is remarkable that the system is more or less recovered to the original equilibrium state before the irradiation and the light absorbs most of the energy at the time of decreasing laser intensity.

In addition, after the laser irradiation, the superconducting long-range order still has the Higgs oscillation which has been mainly discussed in BCS and phonon-mediated superconductors (*27-31*). We note a caution whether the oscillations in our calculations are Higgs or Rabi oscillations. Rabi oscillation happens only during the laser irradiation and when the frequency of the laser is close to the energy difference between two nondegenerate eigenstates of non-perturbative Hamiltonian. In our calculations, we set the frequency of the laser $\omega/t_{\text{hop}} = 15$ where the system does not have the electronic density of states (see Fig. 3). Therefore, the oscillations in our calculations cannot be considered as the Rabi oscillations.

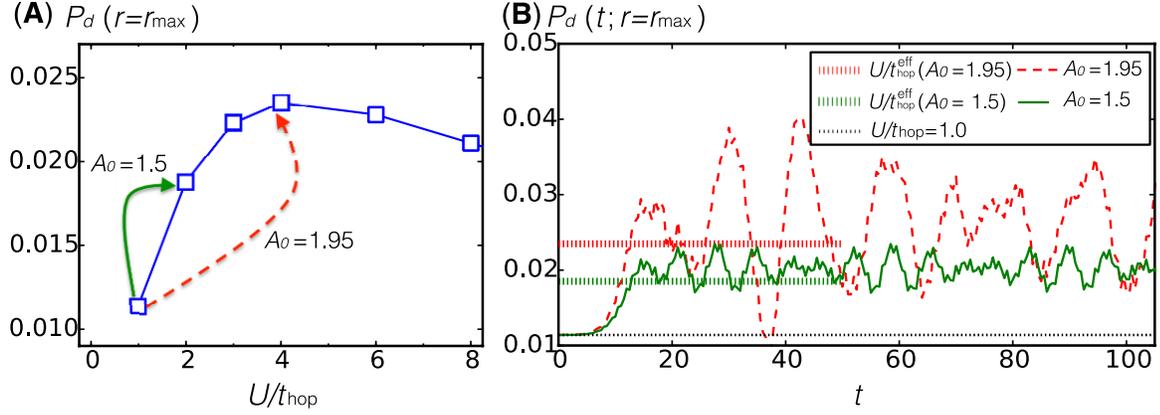

**Fig. S1.**
**Superconducting correlation functions at furthest site $P_d(|r| = r_{max})$ in equilibrum and nonequilibrium obtained from the exact calculations.** (A) Interaction dependence of $P_d(|r| = r_{max})$ for ground states in two-dimensional Hubbard model for $L = 4$ and $\delta = 0.25$. Here, $r_{max} = 2\sqrt{2}$. Red dashed (green solid) arrows indicate the path of the system under the laser irradiations for $A_0 = 1.95$ ($A_0 = 1.5$) and $\omega \gg t_{hop}$. (B) Time evolution of $P_d(|r| = r_{max})$ induced by strong laser in two- dimensional Hubbard model. Parameter set is $(U/t_{hop}, \omega/t_{hop}, t_c) = (1.0, 15, 60)$. Red dashed and Green solid curves represent the results for $A_0 = 1.95$ and $A_0 = 1.5$, respectively. Bold horizontal dotted lines indicate values of superconducting correlation functions at the furthest site for charge-uniform ground states of the corresponding static Hamiltonian, where "corresponding" means the time independent Hamiltonian in the absence of the vector potential and at the modified $t_{hop}$ taken at $t_{hop}^{eff}$. The black thin horizontal dotted line indicates $P_d(|r| = r_{max})$ for static Hamiltonian at $U/t_{hop} = 1.0$.

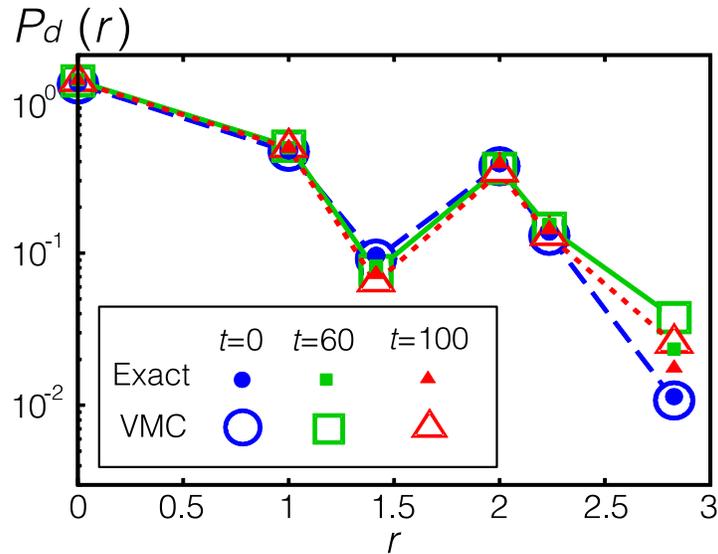

**Fig. S2**
**Dynamics of the superconducting correlations in two-dimensional Hubbard model for $L = 4$ and $\delta = 0.25$ calculated from the exact Schrodinger dynamics**. Solid blue circles, green squares and red triangles represent the results at $t = 0.0, 60.0$ and $100.0$ for $A_0 = 1.95$, respectively. Open symbols represent those of VMC results.

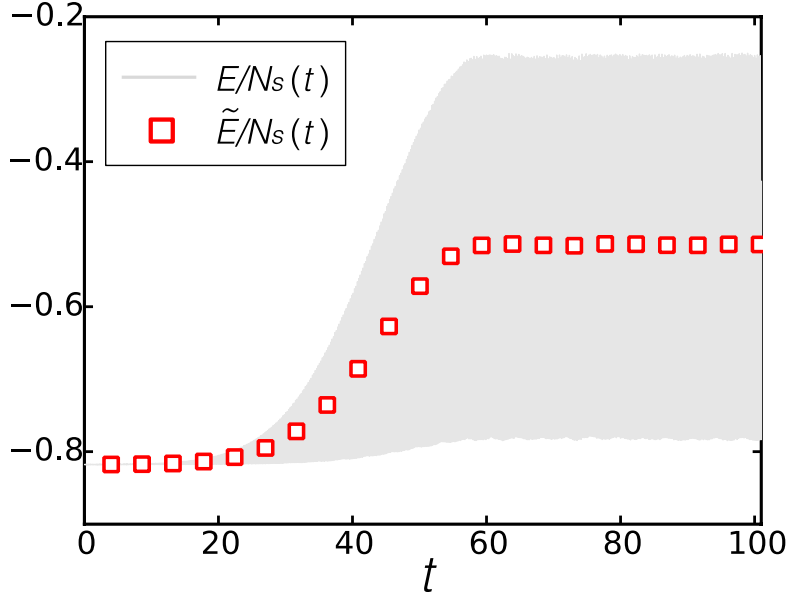

### Fig. S3
**Time-dependence of $E/N_s$ and $\tilde{E}/N_s$ for $L=8$, $\delta=0.0625$ and $U/t_{\text{hop}}=5$.** Grey line and red open square symbols represent the energy per site $E/N_s$ and its average over a AC cycle $\tilde{E}/N_s$, respectively. The parameter set for laser is $(A_0, \omega/t_{\text{hop}}, t_c) = (1.0, 15, 60)$.

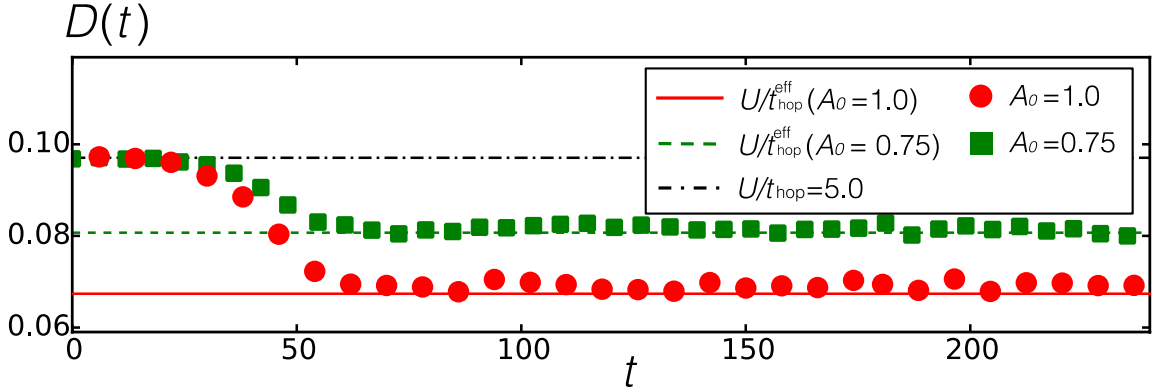

### Fig. S4
**Time-dependence of double occupancy $D$ for $L=8$, $\delta=0.0625$ and $U/t_{hop}=5$.** Green squares and red circles represent the results for $A_0 = 0.75$ and $A_0 = 1.0$, respectively. Red solid and green dashed lines indicate corresponding values of $D$ obtained from VMC for ground states by assuming the effective enhancement of $U/t_{\text{hop}}$. Black dot-dashed line is the equilibrium value in the absence of the laser irradiation. Parameter set for laser is $(\omega/t_{\text{hop}}, t_c) = (15, 60)$.

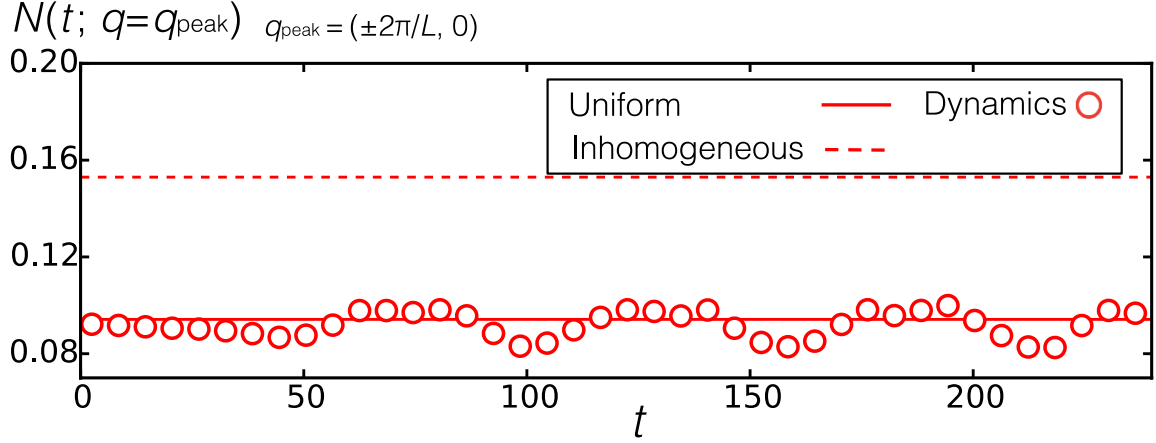

**Fig. S5**
**Time-dependence of peak value of charge structure factor $N(q_{\text{peak}} = (\pm 2\pi/L, 0))$ for $L = 8, \delta = 0.0625$ and $U/t_{hop} = 5$.** Parameter set for laser is $(A_0, \omega/t_{\text{hop}}, t_c) = (1.0, 15, 60)$. Round symbols represent the results for dynamics. Dotted and solid lines indicate the corresponding values of $N(q_{\text{peak}})$ obtained from VMC for ground states with and without charge inhomogeneity by assuming the effective enhancement of $U/t_{\text{hop}}$, respectively.

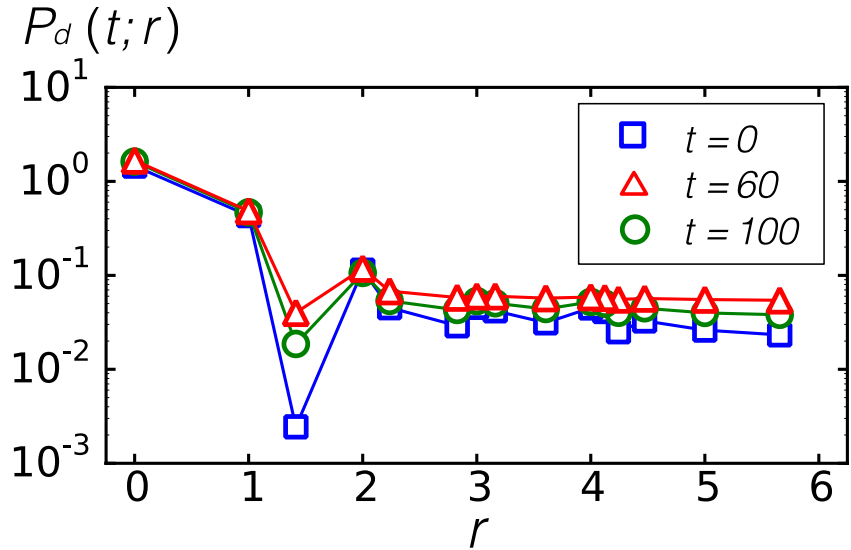

**Fig. S6**
**Time-dependence of superconducting correlations for $L = 8, \delta = 0.0625$ and $U/t_{\text{hop}} = 5$.** Parameter set for laser is $(A_0, \omega/t_{\text{hop}}, t_c) = (1.0, 15, 60)$. Blue squares, red triangles and green circles represent the results at $t = 0.0, 60.0$ and $100.0$, respectively.

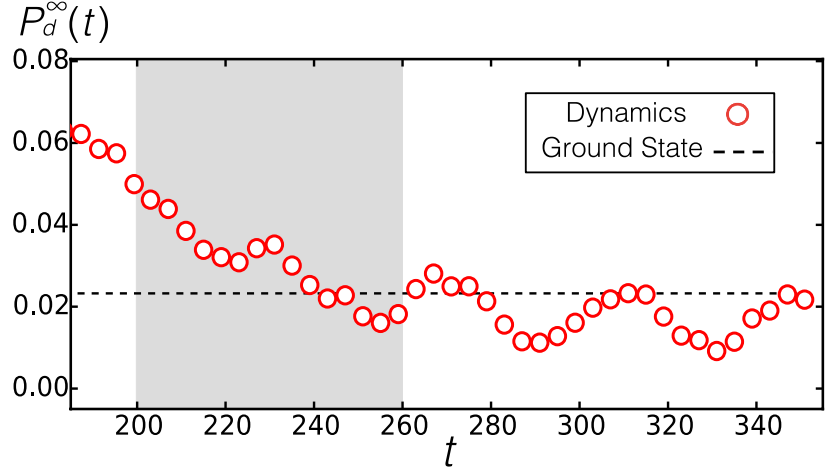

**Fig. S7**

**Time-dependence of $P_d^\infty$ for $L = 8$, $\delta = 0.0625$ and $U/t_{hop} = 5$ during and after the laser irradation.** Red circles represent the results for $A_0 = 1.0$, $\omega/t_{hop} = 15$, $t_c = 60$ and $t_f = 60$. Results before $t = 200$ are the same as those in Fig. 2. Black dashed line indicates the corresponding values of $P_d^\infty$ for the ground state in a system with $U/t_{hop} = 5$ obtained from the VMC calculation. Grey shaded region indicates the falling time region of the laser. After $t = 260$ the vector potential is zero.

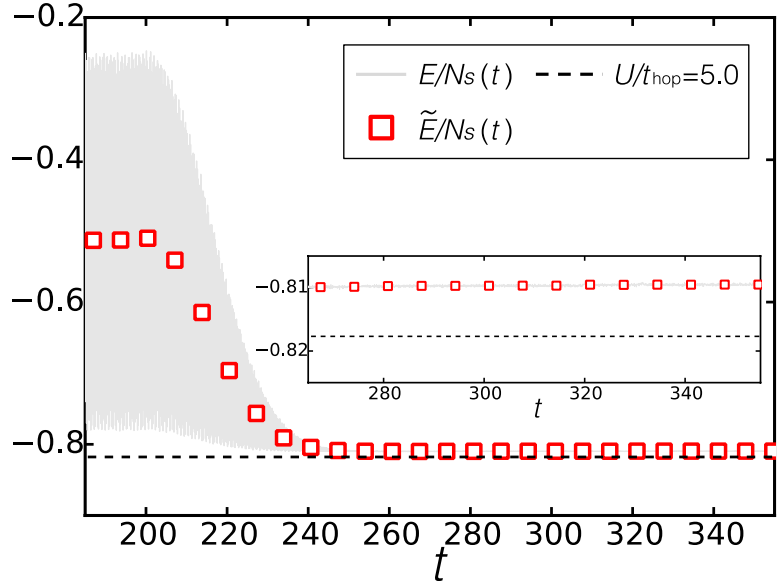

**Fig. S8**

**Time-dependence of $E/N_s$ and $\tilde{E}/N_s$ for $L = 8$, $\delta \approx 0.06$ and $U/t_{hop} = 5$ during and after the laser irradiation.** Red squares and grey thin lines represent the dynamics of $\tilde{E}/N_s$ and $E/N_s$ for $A_0 = 1.0$, $\omega/t_{hop} = 15$, $t_c = 60$ and $t_f = 60$, respectively. Black dashed line indicates the corresponding value of $E/N_s$ for the ground state in a system with $U/t_{hop} = 5$ at equilibrium obtained from the VMC calculation. Inset represents an enlarged view after the laser irradiation.

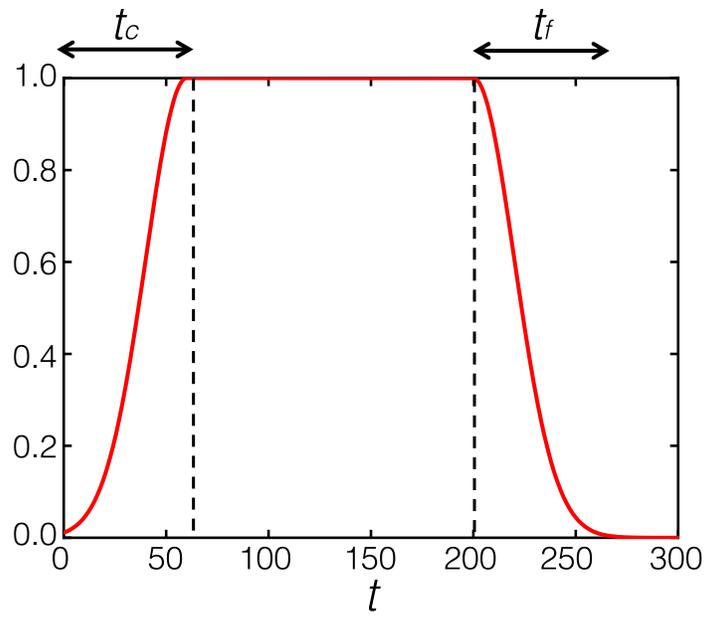

**Fig. S9**
**Envelop function of the laser $g(t)$.** Top left and top right double-headed arrows represents the rising and falling time region, respectively. Black dashed lines indicate the rise end time and the fall start time.